\documentclass[9pt,shortpaper,twoside,web]{ieeecolor}

\usepackage{generic}
\usepackage{cite}
\usepackage{amsmath,amssymb,amsfonts}
\usepackage{graphicx}
\usepackage{textcomp}

\usepackage{amssymb}
\usepackage{latexsym}

\usepackage{url}
\usepackage{xcolor}
\usepackage{hyperref}

\usepackage{multirow,setspace,verbatim,amsfonts,amsmath,amsbsy,epsfig,url}
\usepackage{array}
\usepackage{framed}
\usepackage{kotex}
\usepackage{lineno}
\usepackage{graphicx}
\usepackage{gensymb}
\usepackage{booktabs}
\usepackage{array}
\usepackage{color}
\usepackage{dsfont}

\usepackage{cite}
\usepackage{makecell}
\usepackage{epstopdf}

\usepackage{algorithm}
\usepackage[noend]{algpseudocode}
\usepackage{tikz}
\usepackage{mathrsfs} 
\usepackage[algo2e]{algorithm2e} 

\newcommand{\add}[1] {\textcolor{black}{#1}} 
\usepackage{xcolor,colortbl}

\definecolor{Gray}{gray}{0.85}
\definecolor{LightCyan}{rgb}{0.88,1,1}
\newcolumntype{a}{>{\columncolor{Gray}}c}
\usepackage{marginnote}

\def\BibTeX{{\rm B\kern-.05em{\sc i\kern-.025em b}\kern-.08em
		T\kern-.1667em\lower.7ex\hbox{E}\kern-.125emX}}
\begin{document}
%
%
%
\title{Multi-scale Hybrid Vision Transformer for Learning Gastric Histology: AI-based Decision Support System for Gastric Cancer Treatment}

\author{Yujin Oh, Go Eun Bae, Kyung-Hee Kim, Min-Kyung Yeo$^*$ and Jong Chul Ye$^*$, \IEEEmembership{Fellow, IEEE}
\thanks{This work was supported in part by the Na- tional Research Foundation (NRF) of Korea under Grant NRF-2020R1 A2B5B03001980, in part by the Institute of Information \& communications Technology Planning \& Evaluation (IITP) grant, funded by the Korea government (MSIT) under Grant 2019-0-00075, in part by Artifi- cial Intelligence Graduate School Program (KAIST)), in part by Korea Health Technology R\&D Project through the Korea Health Industry De- velopment Institute (KHIDI), funded by the Ministry of Health \& Welfare, Republic of Korea under Grant HR20C0025, and in part by KAIST Key Research Institute (Interdisciplinary Research Group) Project. (Corresponding authors: Min-Kyung Yeo; Jong Chul Ye.)}
\thanks{Yujin Oh and Jong Chul Ye are with Kim Jaechul Graduate School of Artificial Intelligence, KAIST, 291 Daehak-ro, Yuseong-gu, Daejeon, 34141, Republic of Korea.  (e-mail: yujin.oh@kaist.ac.kr; jong.ye@kaist.ac.kr)}
\thanks{Go Eun Bae and Min-Kyung Yeo are with Department of Pathology, Chungnam National University School of Medicine, Chungnam National University Hospital, Munwha-ro 282, Daejeon, 35015, Republic of Korea.  (e-mail: goeunbae1@gmail.com; mkyeo83@cnu.ac.kr)}
\thanks{Kyung-Hee Kim is with Department of Pathology, Chungnam National University School of Medicine, Chungnam National University Sejong Hospital, 20 Bodeum 7-Ro, Sejong, 30099, Republic of Korea. (e-mail: phone330@cnu.ac.kr)}}

\maketitle

\begin{abstract}
	
Gastric endoscopic screening is an effective way to decide appropriate gastric cancer treatment at an early stage, reducing gastric cancer-associated mortality rate. Although artificial intelligence has brought a great promise to assist pathologist to screen digitalized endoscopic biopsies, existing artificial intelligence systems are limited to be utilized in planning gastric cancer treatment. 
We propose a practical artificial intelligence-based decision support system that enables five subclassifications of gastric cancer pathology, which can be directly matched to general gastric cancer treatment guidance. The proposed framework is designed to efficiently differentiate multi-classes of gastric cancer through multiscale self-attention mechanism using 2-stage hybrid vision transformer networks, by mimicking the way how human pathologists understand histology. 
The proposed system demonstrates its reliable diagnostic performance by achieving class-average sensitivity of above 0.85 for multicentric cohort tests. Moreover, the proposed system demonstrates its great generalization capability on gastrointestinal track organ cancer by achieving the best class-average sensitivity among contemporary networks. Furthermore, in the observational study, artificial intelligence-assisted pathologists show significantly improved diagnostic sensitivity within saved screening time compared to human pathologists.
Our results demonstrate that the proposed artificial intelligence system has a great potential for providing presumptive pathologic opinion and supporting decision of appropriate gastric cancer treatment in practical clinical settings.

\end{abstract}




%

%
%


\begin{centering}
	\begin{figure*}[hbt!]
		\centering
		\centerline{\includegraphics[width=1.0\linewidth]{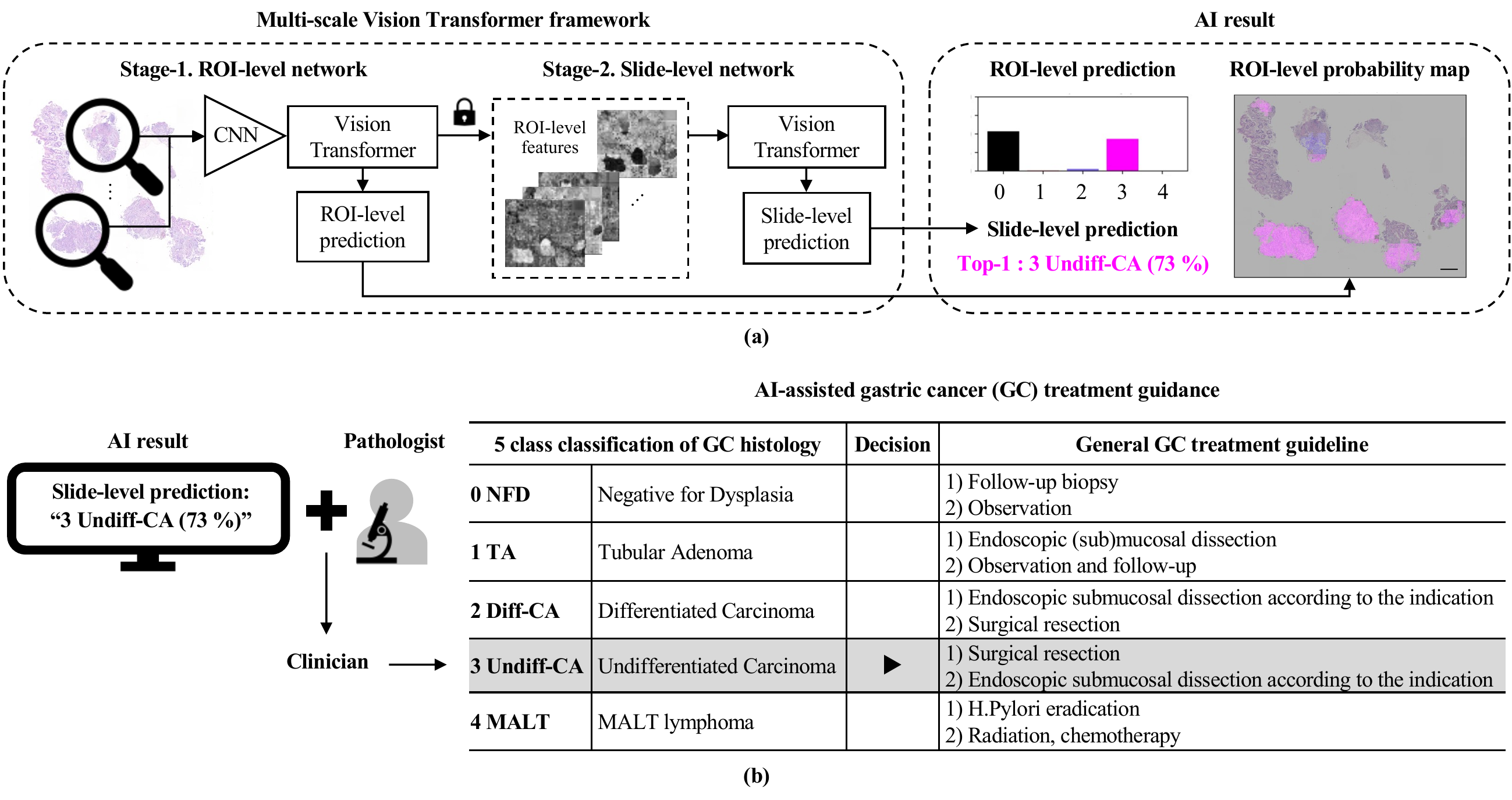}}
		\caption{Schematic of AI-assisted gastric cancer treatment guidance. {\bf (a)} The AI system is composed of 2-stage Vision Transformer modules for comprehensive understanding of ROI-level and slide-level features. {\bf (b)} Once a slide is diagnosed by the AI-assisted pathologist, the clinician can decide gastric cancer treatment based on the five subclassifications result. 
		}
		\label{fig:scenario}
	\end{figure*}
\end{centering}

\section{Introduction}

Gastric cancer (GC) is reported as the fourth most leading cause of cancer death worldwide \cite{sung2021global}. Endoscopic screening is an effective way to detect GC at an early stage, guiding patients to get appropriate treatment according to their cancer stage. 
Although gastric endoscopic screening and proper early treatment have reduced GC-associated mortality rate \cite{jun2017effectiveness}, 
the increasing number of daily endoscopic biopsy cases adds a diagnostic workload to limited clinical resources.
Accordingly, emerging application of artificial intelligence (AI) in the field of digital pathology has brought a great opportunity to effectively reduce diagnostic overloads, by automatically classifying massive number of digitalized whole slide images (WSI). 

AI applications in the field of digital pathology have already achieved powerful diagnostic performance in prostate or breast cancer screening \cite{jiang2020emerging, bulten2020automated, bejnordi2017diagnostic}. AI-assistance systems have also been developed for GC \cite{yoshida2021requirements, song2020clinically, yoshida2018automated, park2021prospective, iizuka2020deep, jang2021deep}. 
An ideal fine-grained classification criteria of GC is presented by World Health Organization (WHO) \cite{nagtegaal20202019}, however, traditional AI systems mostly focus on detecting malignancy over benign or diagnosing only three subclassifications, i.e., benign, adenoma and carcinoma, which cannot fully cover detailed GC subclassifications.
%
Indeed, differentiating fine-grained cancer subclassifications can be challenging even for pathologists, since adjacent diagnostic classes have region-level morphological similarities.  Histologic GC classification needs comprehensive understanding of cell-level to tissue-level morphological features. In particular, when to diagnose a case 
confused between two adjacent classes, pathologists review slides by switching magnification from low to high resolution to understand both global structural relationships and regional morphological features. 

Most existing AI systems for classifying GC employ convolutional neural network (CNN) \cite{song2020clinically, yoshida2018automated, park2021prospective, iizuka2020deep, jang2021deep}.
In order to diagnose gigapixel-level WSI using traditional CNN, patch-level training has become widespread by dividing WSI into sub-patches \cite{liu2017detecting}. However, the patch-wise training is not effective in exploiting relationships between non-adjacent patches and understanding inter-patch structural relationships. To better understand global information, Park and colleagues proposed a RACNN, which aggregates patch-level features using additional convolution layers \cite{park2021prospective}. However, due to limited receptive field size of the convolutional kernel, the additional convolution layers still limit comprehensive understanding of non-adjacent inter-patch relationships.

In contrast, our  in this paper are multifold.
First, inspired by recent success of Vision Transformer (ViT) \cite{dosovitskiy2020image} that exploits long-range dependency between non-adjacent patches through multi-head self-attention mechanism, here we propose a patch-stacked hybrid ViT framework that can significantly expand receptive field, which better understand inter-patch relationship. 
Specifically, as illustrated in Fig. \ref{fig:scenario}(a), our system is composed of 2-stage ViTs: the first stage hybrid ViT encodes region-of-interest (ROI)-level histologic features from stacked patches, which is followed by the second stage ViT for understanding slide-level information.
Specifically, multiple patch stacks from each WSI are fed into the ROI-level network to be trained to match their corresponding patch-stacked annotations. The ROI-level network inference results are then fed into the slide-level network and trained to diagnose its corresponding subclass. 
{As a whole, we can 
 expand receptive field up to 
 the entire size of WSI for comprehensive prediction.}
In this way, the proposed AI system can mimic the entire process of how pathologists understand WSI.

Furthermore, our AI system can classify gastric endoscopic biopsies into 5 categories: negative for dysplasia (NFD), tubular adenoma (TA), differentiated carcinoma (Diff-CA), undifferentiated carcinoma (Undiff-CA), and MALT lymphoma (MALT). As shown in Fig. \ref{fig:scenario}(b), the proposed subclassifications can be matched to the general GC treatment guidance \cite{japanese2011japanese}, thus the results can be directly utilized to guide proper GC treatment in clinical settings. 

To the best of our knowledge, diagnosing five subclasses of GC, including MALT lymphoma, is firstly tried by the proposed AI system.
\add{Specifically, MALT lymphoma has a relatively high prevalence in East Asia related with high Helicobacter pylori (H. pylori) infection \cite{asenjo2007prevalence}.   In South Korea, 1-2\% of patients receiving upper endoscopic biopsy are classified as MALT lymphoma, and it corresponds to 12\% of patients who diagnosed as gastric malignancy \cite{yang2016management}. 
Diagnosis of MALT lymphoma largely depends on the pathologic confirmation, thus classification of MALT class should be listed on the GC screening program, especially in countries with high infection rates of H. pylori. 
Histologic features of MALT lymphoma help to diagnose the disease; however, diagnosis of MALT lymphoma is a challenge even for the pathologist and requires ancillary test, such as immunohistochemistry (IHC) or molecular evaluation, due to its morphological similarities with other inflammatory or tumorous diseases \cite{bacon2007mucosa}.}

\add{Morevoer, the proposed classification additionally categories GC into differentiated and undifferentiated carcinoma. 
Undifferentiated-type carcinoma is reported to have high incidence of lymph node metastasis \cite{gotoda2000incidence, kook2019risk}, thus, identifying undifferentiated carcinoma is important for deciding surgical treatment \cite{gotoda2006endoscopic}. A recent work reported sequential application of a differentiated/undifferentiated binary classifier on a normal/tumor classifier result \cite{jang2021deep}. However, relatively poor performance of undifferentiated carcinoma class indicates the difficulty of discriminating confusing differentiated and undifferentiated cancer cells. }


\add{The proposed AI system demonstrates promising GC classification performance including MALT lymphoma, differentiated and differentiated carcinoma}
by achieving average diagnostic sensitivity of above 0.85 for both internal and external cohort test set. Moreover, the proposed system is proven to have great generalizability on gastrointestinal organ test by achieving the best class-average sensitivity compared to its counterpart networks. Furthermore, in the observation study, pathologists assisted by the AI system show significantly improved diagnostic performance by achieving class-average sensitivity of 0.93 $\pm$ 0.06  compared to human pathologist performance of 0.83 $\pm$ 0.03. The reliable performance of the proposed system and the observer study result demonstrates that the proposed AI system holds great promise in providing practical opinion for guiding appropriate treatment for early-stage GC patients.

\add{Innovation points of our work can be summarized as:}
\add{
\begin{itemize} 
	\item Our work firstly proposes  five subclassification system of GC histology, which can guide treatment in clinical settings
	\item We propose 2-stage multi-scale hybrid vision transformer network, which enables comprehensive histology feature analysis
	\item Our AI system demonstrates its reliable class-average sensitivity above 0.85 on muticentric cohort sets, and assists pathologists to achieve improved prediction sensitivity by 0.10 under improved confidence level and within reduced screening time.
\end{itemize}}

\begin{figure*}[h!]
	\centering
	\centerline{\includegraphics[width=1.0\linewidth]{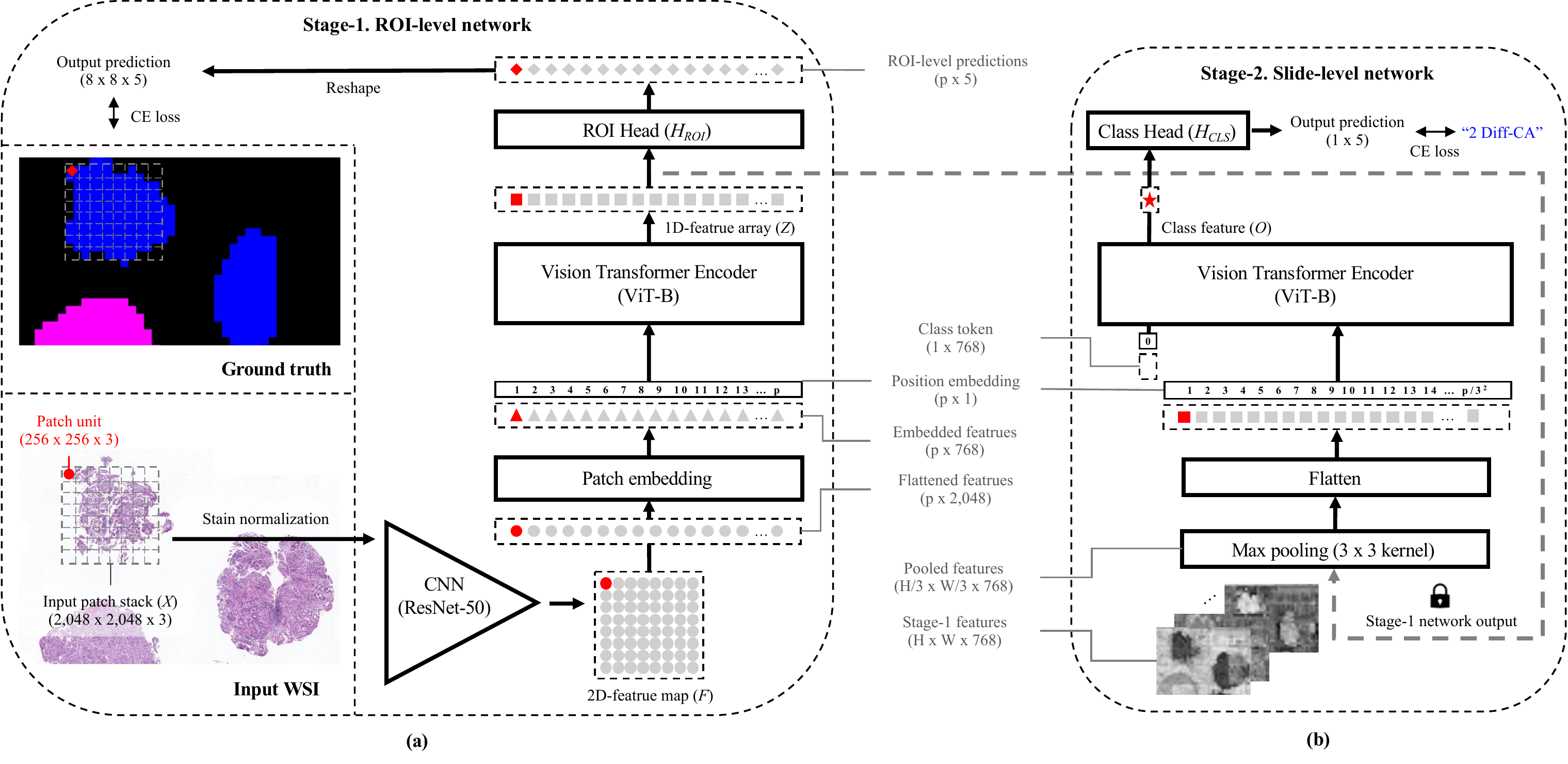}} 
	\caption{\footnotesize {\bf Proposed 2-stage multiscale hybrid Vision Transformer framework.}
		{\bf (a)} ROI-level prediction network with a hybrid Vision Transformer backbone.
		{\bf (b)} Slide-level prediction network with a Vision Transformer backbone.}
	\label{fig:network}
\end{figure*}


\section{Methods}

{Vision Transformer (ViT) has shown state-of-the-art (SOTA) performance in various computer vision tasks \cite{khan2021transformers}. 
By extending this idea,  one of the important contributions of this paper is a 2-stage ViT architecture that effectively learns inter-patch relationships within gigapixel-level WSI through multiscale self-attention mechanism.}
As depicted in Fig. \ref{fig:network}, the AI system is composed of ROI-level and slide-level networks. The ROI-level hybrid ViT network is composed of a convolutional neural network (CNN) module, a ViT encoder module and a ROI-level prediction head, which learn patch-level features from stacked patches. The slide-level ViT network is composed of a ViT encoder module and a slide-level classification head, which learn comprehensive information from stage-1 network output features over the entire slide.
More details are as follows.

\subsection{Stage 1. ROI-level Network}

As illustrated in Fig. \ref{fig:network}(a), for ROI-level prediction, a 2,048 $\times$ 2,048-pixel input patch stack $X$ is firstly stain normalized and fed into the CNN module. The CNN module extracts low-level features given the patch stack $X$, resulting an 8 $\times$ 8 grid 2-D feature map $F$: 
\begin{gather}
	F = \left [
	\begin{array}{ccc}        
		f_1 & \cdots& f_w\\
		\vdots & \ddots & \vdots\\
		f_h & \cdots & f_p \\
	\end{array}
	\right ]
	= CNN \left ( \left [
	\begin{array}{ccc}        
		x_1 & \cdots& x_w\\
		\vdots & \ddots & \vdots\\
		x_h & \cdots & x_p \\
	\end{array}
	\right ] \right ),
\end{gather}
where $f_p$ denotes a $p$-th feature map and $x_p$ denotes a $p$-th input patch in the patch stack $X$.

The extracted features over the entire patch stack $F$ are then flattened and embedded. The embedded patches added by positional embeddings are then fed into the ViT encoder module composed of successive self-attention layers to output an encoded feature array $Z$:
\begin{gather}
	Z = \left [z_1, z_2, ..., z_p \right ],
\end{gather}
where $z_p$ denotes a $p$-th encoded feature in $Z$.

The encoded features $Z$ are then linearly projected through the ROI-level prediction head and reshaped into its original 2D-feature map dimension. The ROI-level output predictions are trained to be matched with their corresponding patch-level ground truths by minimizing standard cross entropy (CE) loss:
\begin{gather}
	\mathcal{L}_{CE_{stage-1}} = -\sum_{cls \in C} \sum_{p \in P} \mathds{1}(y_{p} = cls){\log(H_{ROI}(z_p))}, 
\end{gather}
where $z_p$ denotes a $p$-th encoded feature in $Z$, $H_{ROI}$ denotes the softmax probability of the ROI-level prediction head output, $y_p$ denotes the ground truth label for a $p$-th patch, $P$ denotes the set of patches in the patch stack, $\mathds{1}(\cdot)$ denotes the indicator function,  and $C$ denotes the set of GC subclassification.

\subsection{Stage 2. Slide-level Network}

As depicted in Fig. \ref{fig:network}(b), the slide-level network takes intermediate features $Z$ inferenced by the stage-1 network over  entire receptive field of a slide. To effectively cover the entire receptive field of the gigapixel-level WSI within limited patch grids of the ViT module, the stage-2 network adapted a max pooling module translating 3 $\times$ 3 adjacent patches into a representative 1 $\times$ 1 feature. As a whole, the entire slide-level receptive field can be expanded up to a 96 $\times$ 96 grid 2-D feature map, which correspond to a 12 $\times$ 12 mm slide. 

The pooled input features are then flattened to be fed into the ViT encoder (ViT-B) module. In front of the flattened feature array, a learnable class token with identical feature dimension is prepended, which can attend to the entire feature embeddings throughout successive self-attention layers. The encoded class feature $O$ is then linearly projected to be matched to its corresponding slide-level ground truth by minimizing class-weighted CE loss: 
\begin{gather}
	\mathcal{L}_{CE_{stage-2}} = -\sum_{cls \in C} w_{CLS} \mathds{1}(y = cls){\log(H_{CLS}(O))}, 
\end{gather}
where $O$ denotes the encoded class feature, $H_{CLS}$ denotes the softmax probability of the slide-level classification head output, $y$ denotes the slide-level ground truth label, and $w_{CLS}$ denotes a weight for adjusting unbalanced class distribution.



\subsection{Model Implementation}
\label{imple}

For training the ROI-level network, we utilized ResNet50 \cite{he2016deep} and ViT-B \cite{dosovitskiy2020image} for each CNN and ViT backbone, respectively. The batch size was set 4 and the model was trained for 30 epochs with the initial learning rate of 0.00001. For optimizing the ROI-level network training, we used Adam with decoupled weight decay (AdamW) optimizer with cosine scheduler for learning rate and weight decay. We applied geometric augmentation including random rotating, flipping and scaling with additional data augmentation techniques following BYOL \cite{grill2020bootstrap} (color jittering, gaussian blurring and solarization). 

For training the slide-level network, we utilized ViT-B \cite{dosovitskiy2020image} as backbone. The batch size was set 5 and the model was trained for 50 epochs with initial learning rate of 0.0004. For optimizing the slide-level network training, we used the stochastic gradient descent (SGC) optimizer with cosine scheduler.  For providing clinician-friendly results, the slide-level output was converted to probability score ranging from 0 to 100\% by using Softmax function. 

All the experiments were performed using Python version 3.9 and Pytorch library version 1.10 on a Nvidia RTX 3090 GPU. 



\section{Experiments}

\subsection{Data Annotation}
	
For training the proposed model, hematoxylin and eosin (H\&E) stained slides were collected from the archives of Chungnam National University Hospital (CNUH), and were scanned with a Pannoramic 250 (3DHISTECH) scanner at $\times$40 magnification. For the internal training dataset, an expert pathologist annotated ROI-level labels into five subclasses using ImageJ \cite{abramoff2004image}. Cases with negative for any tumorous condition and background pixels were classified as NFD. Cases with low to high grade dysplasia were classified as TA. Malignant tumor was divided into Diff-CA, Undiff-CA, and MALT lymphoma. Diff-CA and Undiff-CA were classified by following Japanese classification guideline that was suggested for endoscopic resection \cite{japanese2011japanese}. Diff-CA included well to moderately differentiated tubular/papillary adenocarcinoma, and Undiff-CA included poorly differentiated tubular/poorly cohesive/signet ring cell (SRC)/mucinous adenocarcinomas. MALT class was diagnosed followed by Wotherspoon criteria, which was score 4 or above \cite{wotherspoon1993regression}. 
\add{For all the datasets, the slide-level ground truth was established by three experienced gastrointestinal pathologists. A lead expert pathologist primarily classified histology to five subclasses, and the ground truth was confirmed by two other pathologists. For difficult cases, the ground truth was established through consensus between three pathologists.}

\subsection{Datasets}

\begin{centering}
	\begin{figure}[!h]
		\centering
		\centerline{\includegraphics[width=0.98\linewidth]{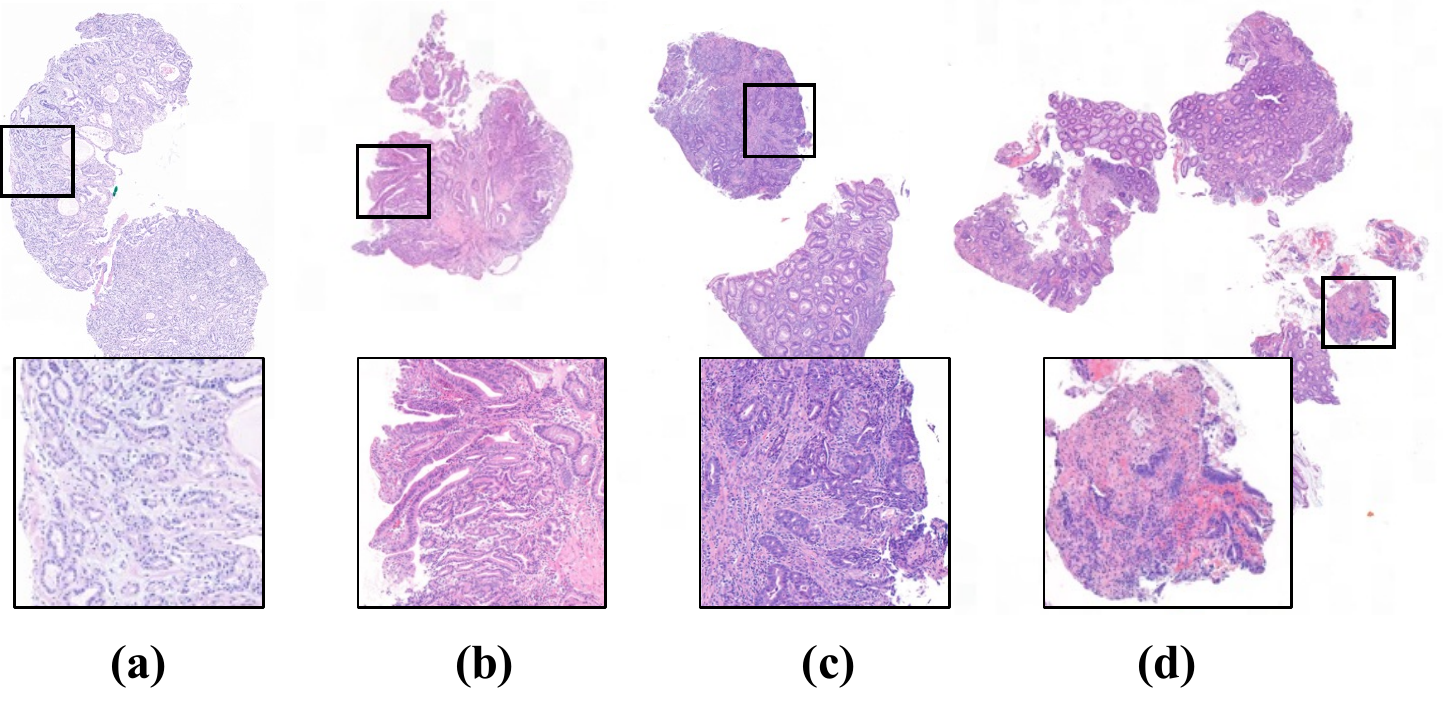}}  
		\caption{Representative slides from Diff-CA class selected from different dataset: 
			{\bf (a)} internal stomach train set, {\bf (b)} internal stomach cohort set, {\bf (c)} external stomach cohort set and {\bf (d)} external colon cohort set.}
		\label{fig:sample}
	\end{figure}
\end{centering}

\subsubsection{Internal Stomach Dataset}
	
The internal training dataset were collected from CNUH, including 1,228 endoscopic biopsy slides. 
The internal stomach dataset were randomly split into train, validation and internal test set of 70, 10 and 20 percentiles from the entire dataset, respectively, following class distribution of the entire dataset. A detailed dataset distribution is described in Table \ref{Table:dataset}. 

For the convenience of training the AI system, one clear WSI per each slide was collected automatically and down-sampled at $\times$20 magnification with resolution of 0.485 $\mu m$ per pixel. 2,048$\times$2,048-pixel patch stacks were sampled from all the train set. 
 \add{From each WSI, patch stacks were collected by sliding window which allow limited overlapped pixels between each patch stack, and  for each slide, we collected at most 100 patch stacks.} 
Patch stacks containing foreground tissues below 30 percentiles to the entire patch stack size were excluded from the train set. 

\subsubsection{Internal Stomach Cohort Set}

The internal stomach cohort set was collected from CNUH, including 876 slides from endoscopic biopsy slides cohort from June 2021 to July 2021. 

\subsubsection{External Stomach \& Colon Cohort Set}

The external cohort sets were prepared for both stomach and colon specimens, which were collected from Sejong Hospital of Chungnam National University Hospital (CNUSH), including 336 and 400 slides respectively. These datasets have different slicing and staining characteristics to that of the training dataset, as shown in Fig. \ref{fig:sample}. The specimens were collected from endoscopic biopsy slides cohort from September 2020 to February 2021.

\begin{centering}
	\begin{table}[!ht] \caption{\bf\footnotesize Class distribution of training and test dataset. }  \label{Table:dataset}
		\centering
		\scalebox{0.77}{
			\begin{tabular}{l|cccc|cc|cc}
				
				\hline 
				~ & \multicolumn{4}{c|}{\textbf{Int. train se}t}  & \multicolumn{2}{c|}{\textbf{Int. val set}}  & \multicolumn{2}{c}{\textbf{Int. test set}}  \\  
				\textbf{Class} & \#slides & \% & \#PS. & \%  &  \#slides &  \%    &  \#slides &  \%   \\ \hline 
				\textbf{0 NFD}  & 263 & 31 & 1,850 & 15 & 38 & 28   & 73 & 30  \\ 
				\textbf{1 TA} & 160 & 19 & 1,825 & 15 & 24 & 18 & 43 & 17   \\ 
				\textbf{2 Diff-CA} & 175 & 21 & 2,977 & 24 & 31 & 23  & 54 & 22  \\ 
				\textbf{3 Undiff-CA} & 180 & 21 & 5,032 & 40 & 33 & 24  & 58 & 23  \\ 
				\textbf{4 MALT} & 68 & 8 & 826 & 7 & 9 & 7 & 19 & 8  \\ \hline  
				\textbf{Total} & 846 & ~ & 12,510 & ~ & 135 & ~  & 247 & ~  \\ 
				\hline 
				
				
				~ & \multicolumn{4}{c|}{\textbf{Int. stomach set}}  & \multicolumn{2}{c|}{\textbf{Ext. stomach set}}  & \multicolumn{2}{c}{\textbf{Ext. colon set}}  \\  
				\textbf{Class} & \#slides & \% & \#PS. & \%  &  \#slides &  \%    &  \#slides &  \%    \\ \hline 
				\textbf{0 NFD}  & 772 & 88 & ~ & ~  & 297 & 88 & 221 & 55  \\ 
				\textbf{1 TA}  & 47 & 5 & ~ & ~ & 11 & 3   & 165 & 41  \\ 
				\textbf{2 Diff-CA} & 26 & 3  & ~ & ~  & 12 & 4  & 13 & 3 \\ 
				\textbf{3 Undiff-CA} & 19 & 2 & ~ & ~  & 8 & 2  & 1 & 0 \\ 
				\textbf{4 MALT} & 12 & 1 & ~ & ~ & 8 & 2 & - & - \\ \hline  
				\textbf{Total} & 876 & ~ & ~ & ~ &  336 & ~ &  400 & ~  \\
				\hline 
				\multicolumn{9}{l}{\textit{Note:} Int.: Internal; Ext.: External; PS: Patch stacks} \\
				
				
		\end{tabular}}
	\end{table}
\end{centering}

\begin{centering}
	\begin{figure}[hbt!]
		\centering
		\centerline{\includegraphics[width=1.0\linewidth]{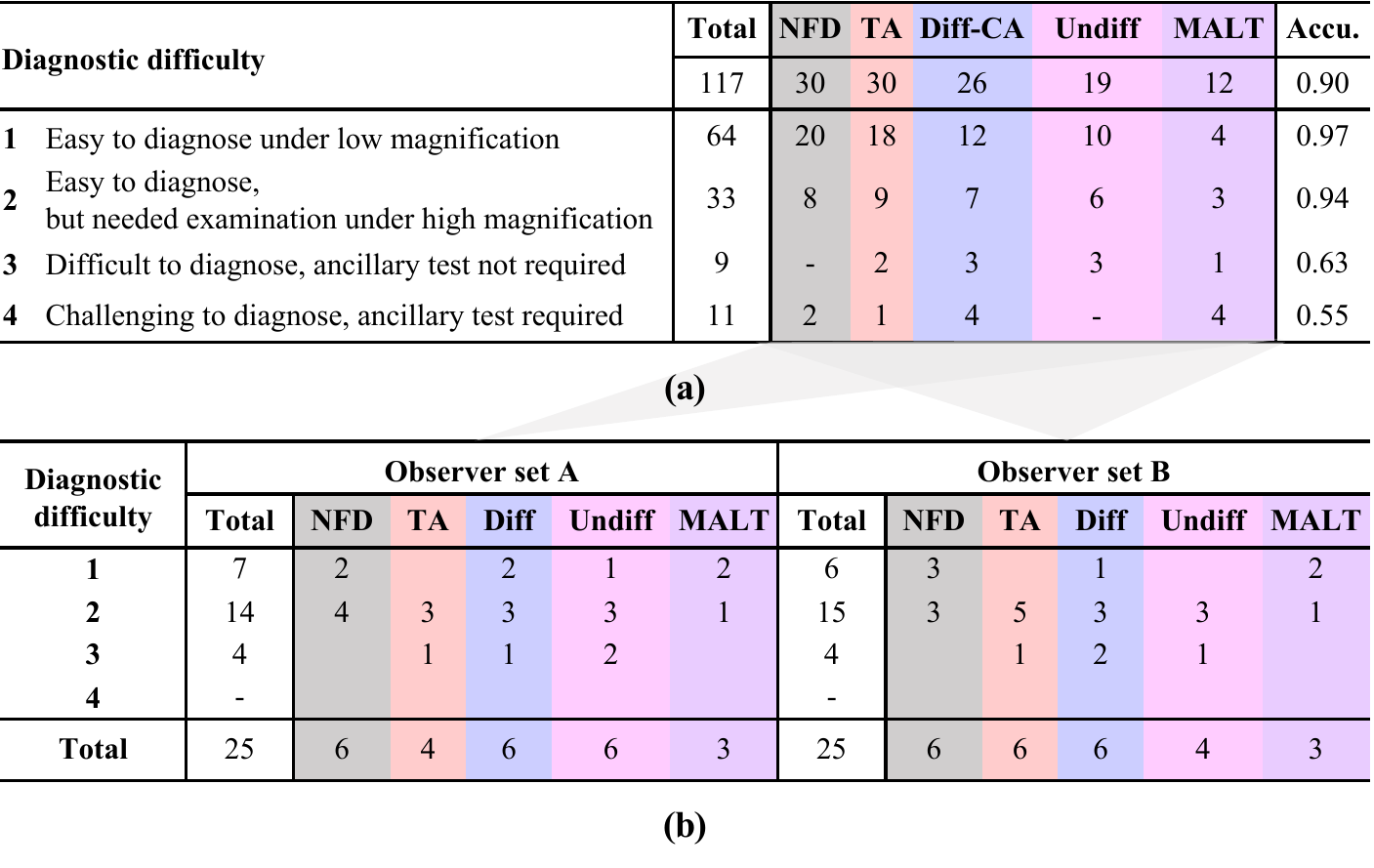}}
		\caption{Observer study design.
			{\bf (a)} Diagnostic difficulty analysis on internal stomach cohort set. 
			{\bf (b)} Observer test sets distribution. }
		\label{fig:observer}
	\end{figure}
\end{centering}

\subsection{Observer Study Design}

The observer study was designed to evaluate the AI-assisted pathologist performance on daily gastric endoscopic screening. As shown in Fig. \ref{fig:observer}, from 117 cases of the sampled internal cohort test set, 2 observer test sets (n = 25/set) were prepared considering the class distribution and the diagnostic difficulty level. 
The diagnostic difficulty was stratified following previous trial \cite{song2020clinically}: level 1, easy to diagnose under low magnification; level 2, easy, but needed examination under high magnification; level 3, difficult, but ancillary test not required; level 4, challenging and ancillary test required. \add{Then, three expert gastrointestinal pathologists established the diagnostic difficulty ground truth for the entire observer test set through a consensus meeting.}
We excluded cases with the highest diagnostic difficulty, for avoiding potential diagnostic disagreement. 

All the participants were given documented AI system performance and classification guidelines justified by an expert pathologist participated in the annotation process. For each test set, participants were given 25-minutes of time constraints. Neither case revision nor answer correction was restricted. Total 6 pathologists from CNUH and CNUSH were participated in the observer test. Among the participants, 4 pathologists of over 10 years of experience were equally split into different test groups. A pathologist of over 5 years of experience and a pathology resident of over 4 years of experience were split into different test groups. 

Participants were asked to diagnose WSI in each observer test set with AI-assistance (AI-assisted) and without AI-assistance (Pathologist-only), respectively. The order of the AI-assistance and the pathologist-only trials was pre-determined, as described in Fig \ref{fig:stat}. Minimum 3-hours of break time was given between each test sets. 
The participants were asked to fill out answers in tables (Excel 2019, Microsoft), and asked to score their confident level of the diagnosis from least (0) to most (1.00) for each case. Confident level of the diagnosis was justified as follows: indefinite for, $<$0.50; suspicious for, 0.51-0.70;  favor, 0.71-0.80; consistent with, 0.81-0.90; diagnostic of, 0.91-1.00, following \cite{lindley2014communicating}. 


\subsection{Evaluation Metrics}
\label{metric}

{For evaluating the ROI-level network performance, we utilized patch-level accuracy per each WSI. The metric is defined as follows:}
\begin{gather*}
	{{Accuracy}_{patch}}(y, \hat{y}) =  \frac{1}{\left|S_\text{c}\right|} \sum_{s \in S_\text{c}} \Bigg( \frac{1}{\left|P_\text{s}\right|} \sum_{p \in P_\text{s}} M_p \cdot 1(\hat{y}_p = y_p) \Bigg), 
\end{gather*}
where $S_\text{c}$ is the set of WSI samples of each slide-level truth class, $P_\text{s}$ is the set of patches within $s$-th WSI sample, $\hat{y}_p$ is the predicted value of the $p$-th patch and $y_p$ is the corresponding patch-level annotation label, and $M_p$ is foreground mask, which is set to $1$ when a patch consists of more than 10\%  of foreground pixels.

For evaluating the slide-level multi-class classification performance, we utilized class-wise slide-level accuracy, specificity and sensitivity, which are defined as follows: 
\begin{gather*}
	{Accuracy} = \frac{tp + tn}{tp + tn + fp + fn}, \\
	{Specificity} =  \frac{tn}{tn + fp}, \\
	{Sensitivity} = \frac{tp}{tp + fn},
\end{gather*}
where ${tn},  {fp}, {tp}, {fn}$ represent true negative, false positive, true positive and false negative cases for each class, respectively. 
All the metrics were by calculated using Scikit-learn package \cite{scikit-learn}.

For averaging multi-class metrics or multi-observer test metrics, we utilized macro average as default. The metric is defined as follows:
\begin{gather*}
	{Average} = \frac{1}{\left|C\right|} \sum_{{cls} \in C}M_{cls},
\end{gather*}
where $C$ is the set of slide-level ground truth classes and $M_{cls}$ represents any class-wise performance metric, e.g., accuracy,  specificity and sensitivity.

For averaging the entire observer test records, e.g., confident level of the diagnosis and screening time, we utilized micro average. The metric is defined as follows:
\begin{gather*}
	{Average (micro)} = \frac{1}{\left|S\right|} \sum_{s \in S} s,
\end{gather*}
where $S$ is the set of observer test records over all the participants.

\subsection{Statistical Analysis}

Statistical analysis for the observer test was performed using MATLAB R2020a (Mathworks, Natick). Kolmogorov Smirnov test was used to evaluate normality of all the results. Since all the test results were non-normally distributed, a two-sided Wilcoxon rank sum test was used to statistically compare performance metrics. P-value was utilized as statistical significance level and indicated as asterisks, i.e., * for p$<^{}$0.05; **  for p$<$0.01. \add{We visualized the statistical effects of the observer test by following a previous work \cite{zhou2020deep}.}

\subsection{Ethical Approval}
This study was approved by the institutional review board of Chungnam National University Hospital (IRB file no. 2021-10-028) and Chungnam National University Sejong Hospital (IRB file no. 2021-12-004), which waived the requirement for informed consent. 

\section{Experimental Results}
\label{sec:result}

\begin{centering}
	\begin{figure}[!t]
		\centering
		\centerline{\includegraphics[width=1\linewidth]{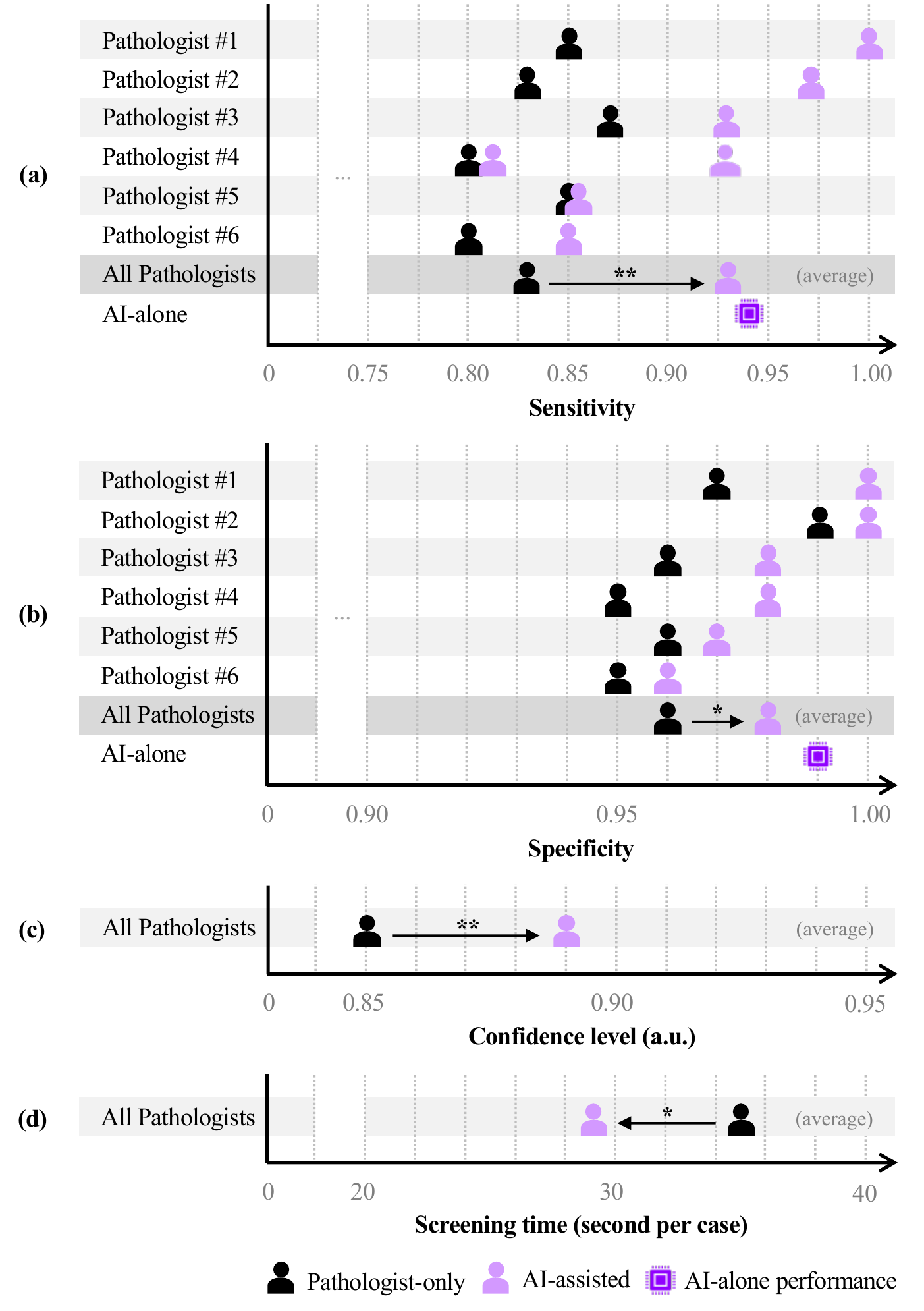}}  
		\caption{Diagnositc performance comparison between pathologist-only, AI-assisted and AI-alone trials on
			{\bf (a)} sensitivity, {\bf (b)} specificity, {\bf (c)} diagnostic confidence level, and {\bf (d)} screening time for each slide.}
		\label{fig:stat}
	\end{figure}
\end{centering}

\subsection{ROI-level Prediction Performance}

We first evaluated whether the proposed patch-stack hybrid vision transformer model better understood inter-patch relationship. As shown in Table \ref{Table:metric}(a), The proposed ROI-level network achieved the patch-level accuracy of 0.85 $\pm$ 0.06, which was the best performance among other baseline networks (see Table \ref{Table:cnn_comparison}). Fig. \ref{fig:cnn_vit} illustrates the representative ROI-level probability map. The proposed ROI-level probability map highly corresponds to the ground truth annotation. The class-wise diagnostic performance showed that our ROI-level network was mostly confused when predicting the Diff-CA class patches between TA or Undiff-CA classes, due to their patch-level morphological similarity.

\begin{centering}
	\begin{table}[!h] \caption{\bf\footnotesize ROI-level network performance comparison on stomach datasets.\\}  \label{Table:cnn_comparison}
		\centering
		\scalebox{0.8}{
			\begin{tabular}{l|ccc|c}
				\hline 
				
				& \multirow{2}{*}{\textbf{Magn.}} & \textbf{Resolution}& \multirow{2}{*}{\shortstack{\textbf{Receptive field }\\ (H  $\times$ W, mm)}} &  \multicolumn{1}{c}{\textbf{Internal test set}} \\
				& ~ & ($\mu m$/pixel) &~ & ${{Accuracy}_{patch}}$ \\  \hline \hline
				{{ResNet50}} &  {$\times$20} &  {0.49} & {0.13 $\times$ 0.13} & {0.79 $\pm$ 0.11}  \\
				\cline{1-5}
				
				InceptionV3 &  {$\times$20} &  {0.49} & {0.13 $\times$ 0.13} & {0.80 $\pm$ 0.11}  \\
				\cline{1-5}
				
				{{ViT}}   &  {$\times$20} &  {0.49} & {0.13 $\times$ 0.13} &  {0.82 $\pm$ 0.10}   \\ 
				\hline 
				
				\multirow{3}{*}{\shortstack[l]{\textbf{Patch-stacked} \\ \textbf{hybrid ViT} \\ \textbf{(Proposed)} }} & \multirow{1}{*}{$\times$20} &  \multirow{1}{*}{0.49} & \multirow{1}{*}{1 $\times$ 1} &  \multirow{1}{*}{\textbf{0.85 $\pm$ 0.06}}  \\   
				
				~ &  $\times$10 & 0.97 & 1.5 $\times$ 1.5 & 0.85 $\pm$ 0.08 \\  
				
				~ &  $\times$5 & 1.95 & 2 $\times$ 2 & 0.84 $\pm$ 0.07 \\	\hline

				\multicolumn{5}{l}{\textit{Note:} Magn.: maginification ratio} \\
				
		\end{tabular}}
	\end{table}
\end{centering}

\subsection{Slide-level Prediction Performance}

We then evaluated slide-level prediction performance for each trial. As shown in Table \ref{Table:metric},
the proposed slide-level network achieved class-average sensitivity of 0.93 $\pm$ 0.05,  0.87 $\pm$ 0.12, and 0.86 $\pm$ 0.13 for each internal test set, internal stomach cohort test, and external stomach cohort test, respectively. The comparable diagnostic sensitivity throughout three different trials demonstrated that the proposed model could be generalized to multicentre dataset with reliable performance. 
We further analyzed \add{detailed classification performance on external stomach test set in Table \ref{Table:detailed_result}}and error cases of each trial in section \ref{section:error}.

\begin{centering}
	\begin{table}[!ht] \caption{\bf\footnotesize Diagnostic performance of the proposed AI system. }  \label{Table:metric}
		\centering
		\scalebox{0.80}{
			\begin{tabular}{l|c|ccc}
				\hline 
				\multirow{2}{*}{\textbf{Class}} & \textbf{ROI-level} & \multicolumn{3}{c}{\textbf{Slide-level}} \\ 
				~ & ${{Accuracy}_{patch}}$ & Accuracy  & Specificity & Sensitivity\\ \hline  \hline
				\textbf{0 NFD} & 0.97 & 0.96 & 1.00 & 0.88  \\ 
				\textbf{1 TA} & 0.82 & 0.97 & 0.98 & 0.93  \\ 
				\textbf{2 Diff-CA} & 0.79 & 0.97 & 0.98 & 0.93   \\ 
				\textbf{3 Undiff-CA}  & 0.87 & 0.97 & 0.96 & 1.00  \\ 
				\textbf{4 MALT}  & 0.83 & 0.98 & 0.99 & 0.89   \\ \hline 
				\textbf{Average} & 0.85 $\pm$ 0.07 &  0.97 $\pm$ 0.01 & 0.98 $\pm$ 0.01 & 0.93 $\pm$ 0.05 \\ 
				\hline 
				\multicolumn{5}{l}{~} \\ 
				\multicolumn{5}{c}{\textbf{(a) Internal test set}} \\ 
				\multicolumn{5}{l}{~} \\ 
				\hline 
				\multirow{2}{*}{\textbf{Class}} & \textbf{ROI-level} & \multicolumn{3}{c}{\textbf{Slide-level}} \\ 
				~ & ${{Accuracy}_{patch}}$ & Accuracy & Specificity & Sensitivity \\ \hline  \hline
				\textbf{0 NFD} & -  & 0.89 & 0.95 & 0.89  \\ 
				\textbf{1 TA} & -  & 0.95  & 0.95 & 0.89  \\ 
				\textbf{2 Diff-CA} & - & 0.98 & 0.97 & 0.88   \\ 
				\textbf{3 Undiff-CA}  & - & 0.97 & 0.97 & 1.00  \\ 
				\textbf{4 MALT}  & - & 0.98  & 0.98 & 0.67   \\ \hline 
				\textbf{Average} & - &  0.95 $\pm$ 0.04 & 0.97 $\pm$ 0.02 & 0.87 $\pm$ 0.12 \\ 
				\hline 
				\multicolumn{5}{l}{~} \\ 
				\multicolumn{5}{c}{\textbf{(b) Internal stomach cohort test}}  \\ 
				\multicolumn{5}{l}{~} \\ 
				
				\hline 
				\multirow{2}{*}{\textbf{Class}} & \textbf{ROI-level} & \multicolumn{3}{c}{\textbf{Slide-level}} \\ 
				~ & ${{Accuracy}_{patch}}$ & Accuracy & Specificity & Sensitivity\\ \hline  \hline
				\textbf{0 NFD} & -  & 0.82 & 0.97 & 0.80 \\ 
				\textbf{1 TA} & -  &  0.95  & 0.95 & 0.73  \\ 
				\textbf{2 Diff-CA} & - & 0.96 & 0.94 & 0.75  \\ 
				\textbf{3 Undiff-CA}  & - & 0.91  & 0.88 & 1.00  \\ 
				\textbf{4 MALT}  & - & 0.97 & 0.96 & 1.00  \\ \hline 
				\textbf{Average} & - &  0.92 $\pm$ 0.06 & 0.96 $\pm$ 0.03 & 0.86 $\pm$ 0.13 \\
				\hline 
				\multicolumn{5}{l}{~} \\ 
				\multicolumn{5}{c}{\textbf{(c) External stomach cohort test}}   
				
		\end{tabular}}
	\end{table}
\end{centering}


\begin{table}[h!] \caption{\bf\footnotesize Detailed classification performance on external stomach cohort test. }  \label{Table:detailed_result}
	\centering
	\scalebox{0.74}{
		\begin{tabular}{l|ccccc} 
			\hline
			\multirow{2}{*}{\textbf{Detailed classification}} & \multicolumn{5}{c}{\textbf{Sensitivity}}   \\ 
			~ &{NFD} & {TA} & {Diff-CA} & {Undiff-CA}  & {MALT}  \\ \hline \hline 
			
			\textbf{0 NFD}  & ~ & ~ & ~ & ~  \\ 
			-  Foveolar hyperplasia &  \underline{1.00} & ~ & ~ & ~ & ~   \\ 
			- Hyperplastic polyp & \underline{0.80} & 0.10 & ~ & 0.10 & ~   \\  
			- Acute gastritis &  \underline{0.69} & 0.05 & 0.05 & 0.14 & 0.07   \\  
			- Atrophic gastritis & \underline{0.88} & 0.04 & 0.02 & 0.06 & ~   \\ \hline  
			\textbf{1 TA} & ~ & ~ & ~ & ~  \\ 
			- TA, Low grade &  0.13 & \underline{0.75} & 0.13 & ~ & ~   \\  
			- TA, High grade & ~ & \underline{0.67} & 0.33 & ~ & ~   \\ \hline 
			\textbf{2 Diff-CA} & ~ & ~ & ~ & ~  \\ 
			- TAC, WD   & ~ & ~ & \underline{1.00} & ~ & ~   \\  
			- TAC, MD   & ~ & 0.10 & \underline{0.70} & 0.20 & ~   \\ \hline 
			\textbf{3 Undiff-CA}  & ~ & ~ & ~ & ~  \\ 
			- TAC, PD   & ~ & ~ & ~ & \underline{1.00} & ~   \\  
			- Poorly cohesive carcinoma   & ~ & ~ & ~ & \underline{1.00} & ~  \\  
			- Signet ring cell carcinoma & ~ & ~ & ~ & ~  \\ 
			\cline{1-6} 
			\textbf{4 MALT}  & ~ & ~ & ~ & ~  \\ 
			- MALT lymphoma  & ~ & ~ & ~ & ~ & \underline{1.00}   \\ \hline  
			\textbf{Others (excluded)} & ~ & ~ & ~ & ~  \\ 
			- Diffuse large B-cell lymphoma  & ~ & ~ & ~ & 1.00 & ~   \\ \hline 
			
			
			\multicolumn{6}{l}{\textit{Note:} \underline{Underline} {\add{indicates} true positive ratio}; TAC: Tubular adenocarcinoma; } \\
			\multicolumn{6}{l}{WD: Well differentiated; MD: Moderately differentiated; PD: Poorly differentiated} \\
			
	\end{tabular}}
\end{table}

\subsection{Comparative Network Analysis}

The proposed ROI-level performance was further compared with baseline networks, i.e., ResNet50, InceptionV3, and ViT, including traditional patch-level prediction methods utilized in comparative studies \cite{jang2021deep, stegmuller2022scorenet}. As shown in Table \ref{Table:cnn_comparison}, the proposed patch-stacked hybrid ViT showed the most promising class-average patch-level accuracy compared to its counterparts with statistically significant level (P-value of $<$0.001 for all the counterparts). 
Moreover, as shown in Fig. \ref{fig:cnn_vit}, the ROI-level probability map provided the least confusing prediction compared to its counterparts.
The proposed framework was further evaluated with various experimental conditions, and showed the best performance with magnification ratio of $\times$20 and ROI-level receptive field of 1 $\times$ 1 mm.

The proposed slide-level performance was compared with baseline classification methods, i.e., Tok-K Mean, Random Forest, and ResNet50, including traditional classification methods utilized in comparative studies \cite{song2020clinically, park2021prospective}. To focus on multi-class classification capability of the AI system in clinical settings, we utilized class-average sensitivity as a primary criterion for model comparison. As shown in Table \ref{Table:vit_comparison}, The proposed combination showed the best class-average performance among almost all combinations of stage-1 and stage-2 networks. In specific, the proposed model showed reliable external validation performance with class-average sensitivity of above 0.86, whereas other methods showed drastic performance degradation due to overfitting on internal training dataset, e.g., the random forest method showed the best performance on internal test, whereas showed the second-worst performance on external cohort test. 


\subsection{Gastrointestinal Organ Generalization Performance}

The proposed network performance was further evaluated on the external colon cohort set. Gastrointestinal track organs basically share the patch-level morphology, however, care should be taken to diagnose slide-level classification, which need comprehensive understandings of inter-tissue relationship. As shown in Table \ref{Table:colon}, the proposed model achieved class-average sensitivity of 0.77 $\pm$ 0.16, which was the best among all the baseline networks with least inter-class variation level. The experimental result demonstrated that the proposed model had great potential for being generalized to unseen gastrointestinal track organs compared to traditional methods, \add{based on comprehensive understanding of gastrointestinal organ cancer characteristics.} 

\begin{centering}
	\begin{table}[!ht] \caption{\bf\footnotesize Model generalization performance on GI track organ.\\}  \label{Table:colon}
		\centering
		\scalebox{0.79}{
			\begin{tabular}{l|c||l|c}
				\hline 
				{\textbf{Stage-1 Network}} & \multirow{2}{*}{\shortstack{\textbf{Receptive field }\\ (H  $\times$ W, mm)}} & {\textbf{Stage-2 Network}} &  \multicolumn{1}{c}{\textbf{External colon set}}   \\   
				
				ROI-level &~ & Slide-level & Sensitivity \\ \hline  \hline
				
				ResNet50 & \multirow{3}{*}{0.13 $\times$ 0.13} & \multirow{3}{*}{ViT}   &  { 0.73 $\pm$  0.19   }   \\   
				\cline{1-1}\cline{4-4}
				InceptionV3 & ~ & ~  &  0.67  $\pm$ 0.28  \\
				\cline{1-1}\cline{4-4}
				ViT & ~ &  ~  &  0.75 $\pm$ 0.18  \\
				\cline{1-4}
				\multirow{4}{*}{\shortstack[l]{\textbf{Patch-stacked} \\ \textbf{hybrid ViT} \\ \textbf{(Proposed)} }}  &  \multirow{4}{*}{1 $\times$ 1}  & {Top-K Mean}  & 0.58 $\pm$  0.47  \\
				~ & ~ & {Random Forest}  &  { 0.59 $\pm$  0.32   } \\
				~ & ~ & {ResNet50}  &  { 0.73 $\pm$  0.34   } \\
				~ & ~ & \textbf {ViT (Proposed)}  & \textbf{    0.77 $\pm$  0.16   } \\ 
				\hline
				\multicolumn{4}{l}{\textit{Note:} GI.: gastrointestinal} \\
		\end{tabular}}
	\end{table}
\end{centering}

\subsection{Error Case Analysis}
\label{section:error}

We counted error cases of each test trial and further analyzed frequent errors. In the internal test, most error cases were counted when distinguishing between adjacent classes (9 of 247 cases) and false-positive cases of NFD (9 of 73 cases). A representative falsely diagnosed case is shown in Fig. \ref{fig:error}(a). The representative result shows complex ROI-level probabilities of multiple classes, i.e., NFD, Diff-CA and Undiff-CA classes, thus the slide-level prediction is uncertain with probability score of around 70\%. 

In the internal cohort test, similar to the internal test results, most error cases were counted when distinguishing between adjacent classes (8 of 104 cases) and false-positive cases of NFD (87 of 772 cases). Errors cases were also counted when discriminating non-adjacent classes between NFD/MALT (3 of 12 cases) and TA/MALT (1 of 47 cases). 
Fig. \ref{fig:error}(b) depicts a representative case which failed to be diagnosed as MALT class. The ROI-level probability map shows small amount of patches diagnosed as MALT class, thus the ROI-level probability for the MALT class is under-estimated. The result indicates that the AI system shows poor performance on MALT cases  that need ancillary test for accurate diagnosis, which will be further discussed in section \ref{sec:lim}.

In the external cohort test, most error cases were counted when distinguishing between adjacent classes (6 of 39 cases) and false-positive cases of NFD (59 of 297 cases), similar to the internal test results. A representative falsely diagnosed case is shown in Fig. \ref{fig:error}(c). The ROI-level probability map shows small portion of patches which falsely-diagnosed as Undiff-CA class, thus the slide-level prediction is uncertain for top-1 prediction with probability score of around 50\%, followed by the ground truth class of around 40\%.

\subsection{Observer Test Result} 

The practical usability of the AI system was further evaluated through observation of AI-assisted pathologist performance. 
\add{We observed that pathologist performance with AI-assistance trails was improved from pathologist-only trials with statistically significant level, as shown in Fig. \ref{fig:stat}, The class-average sensitivity and specificity were increased by 0.10 from 0.83 $\pm$ 0.02 to 0.93 $\pm$ 0.06, and by 0.02 from 0.96 $\pm$ 0.01 to 0.98 $\pm$ 0.01 with AI-assistance, respectively.
We also found that the improved performance was achieved under improved confident level of the diagnosis from 0.85 $\pm$ 0.13 to 0.89 $\pm$ 0.11 with AI-assistance, and within reduced screening time from 34.80 $\pm$ 27.24 (range 13-48) to 28.53 $\pm$ 23.15 (range 16-44) second with AI-assistance. 
Moreover, the stand-alone AI system performance (AI-alone) was comparable to that of AI-assisted pathologist, and even exceeded pathologist-only trials with 0.94 and 0.99 for sensitivity and specificity, respectively.} 

\begin{centering}
	\begin{figure*}[!h]
		\centering
		\centerline{\includegraphics[width=0.9\linewidth]{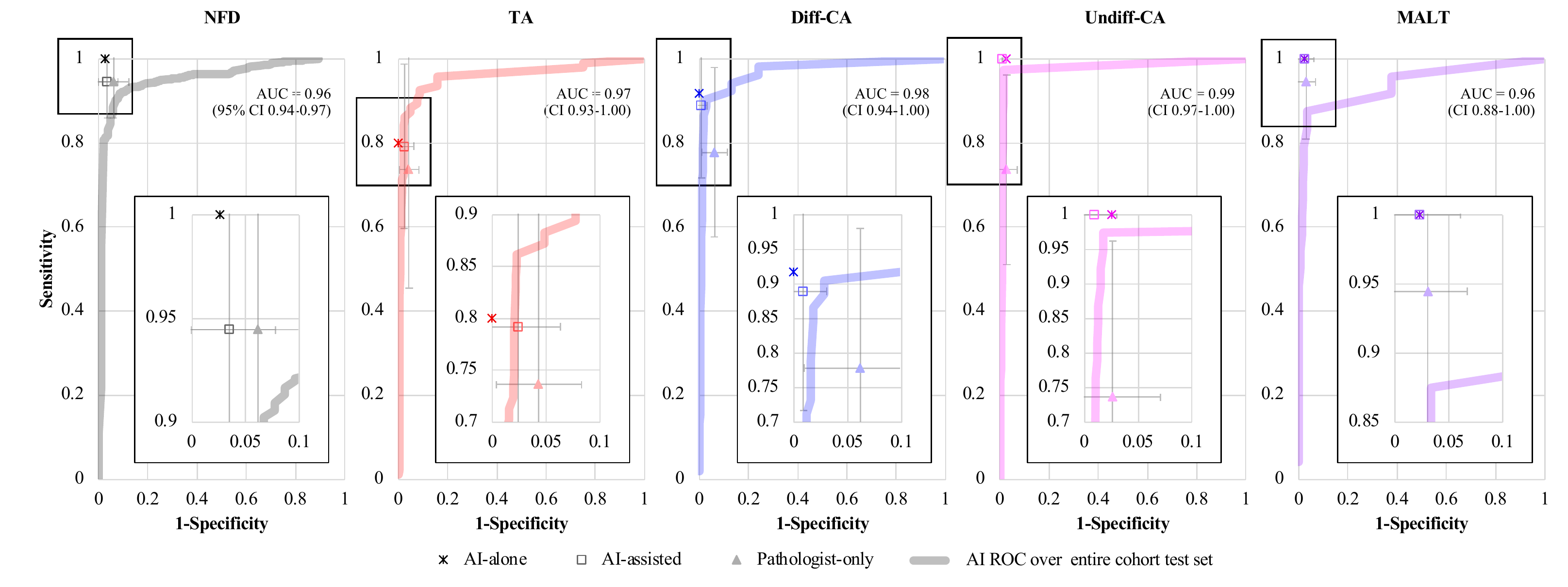}} 
		\caption{Class-wise ROC curve and observer test results. ROC curves represent the proposed AI system performance of the entire internal cohort test set and dot plots represent observer test performance of each AI-alone (asterisk), AI-assisted (square) and pathologist-only (triangle) trials.}
		\label{fig:observer_roc}
	\end{figure*}
\end{centering}

We further evaluated receiver operating characteristics (ROC) curves over the entire internal cohort tests, as shown in Fig. \ref{fig:observer_roc}. AI-assisted pathologist performance for all the classes were improved from pathologist-only performance toward the upper-left side of the curve near the AI-alone performance. We further discovered that the AI system helped pathologists make decision more consistently, by decreasing inter-pathologists variation compared to that of pathologist-only performance, as depicted as error bar of each trial.

\begin{centering}
	\begin{table*}[!h] \caption{\bf\footnotesize Slide-level network performance comparison on vaious stomach datasets.\\}  \label{Table:vit_comparison}
		\centering
		\scalebox{0.885}{
			\begin{tabular}{l|c||l|ccc|ccc|ccc}
				\hline 
				{\textbf{Stage-1 Network}} & {\textbf{Receptive field }} & {\textbf{Stage-2 Network}} &  \multicolumn{3}{c|}{\textbf{Internal test set}} & \multicolumn{3}{c|}{\textbf{Internal cohort set}} & \multicolumn{3}{c}{\textbf{External cohort set}}   \\    
				ROI-level  &  (H  $\times$ W, mm) & Slide-level & Accu. & Spec. & Sens. & Accu.  & Spec. & Sens.  & Accu.  & Spec. & Sens. \\ \hline  \hline
				
				{{ResNet50}}  & {0.13 $\times$ 0.13} & \multirow{3}{*}{ViT (Proposed)} &    0.97    & 0.98 & 0.94 &     0.94    & 0.96 & 0.74 &     0.88    & 0.94 & 0.83  \\
				\cline{1-2} \cline{4-12}
				
				InceptionV3  & {0.13 $\times$ 0.13}   &  ~ &     0.96    & 0.98 & 0.91 &     0.93    & 0.90 & 0.70 &     0.86     & 0.92 & 0.76 \\
				\cline{1-2} \cline{4-12}
				
				{{ViT}}   & {0.13 $\times$ 0.13}  & ~ &     0.97     & 0.98 & 0.92 &     0.93     & 0.94 & 0.68 &    0.84   & 0.91 & 0.78 \\ 
				\hline 
				
				\multirow{6}{*}{\shortstack[l]{\textbf{Patch-stacked} \\ \textbf{hybrid ViT} \\ \textbf{(Proposed)} }}  & \multirow{4}{*}{1 $\times$ 1} & {Top-K Mean}  &     0.83    & 0.88 & 0.49 &    0.97    & 0.87 & 0.41 &     0.97     & 0.86 & 0.47   \\   
				
				~ &  ~ & {Random Forest} &     0.99    & {0.99} & \textbf{0.98} &     0.63    & 0.81 & 0.60 &     0.64    & 0.82 & 0.52  \\
				~ &   ~ & {ResNet50} &     0.96     & 0.98 & 0.92 &      0.96     & 0.94 & 0.68 &    0.96     & 0.93 &  0.78  \\
				~ &   ~ & \textbf{ViT (Proposed)} & {0.97} & {0.99} & {0.97} &  {0.95} & {0.97} & \textbf{0.87} & {0.92} & {0.96} & \textbf{0.86} \\ \cline{2-12}
				
				~ &  1.5 $\times$ 1.5 & \multirow{2}{*}{ViT (Proposed)} &     0.96     & 0.98 & 0.90 &     0.96    & {0.97} & 0.85 &     0.92      & 0.93 & 0.75 \\  \cline{2-2} \cline{4-12}
				
				~ & 2 $\times$ 2 & ~ &     0.97     & 0.93 & 0.92 &     0.94     & 0.95 & 0.82 &    0.91     & 0.93 & 0.80 \\  
				\hline 

		\end{tabular}}

	\end{table*}
\end{centering}

\begin{centering}
	\begin{figure*}[!h]
		\centering
		\centerline{\includegraphics[width=0.9\linewidth]{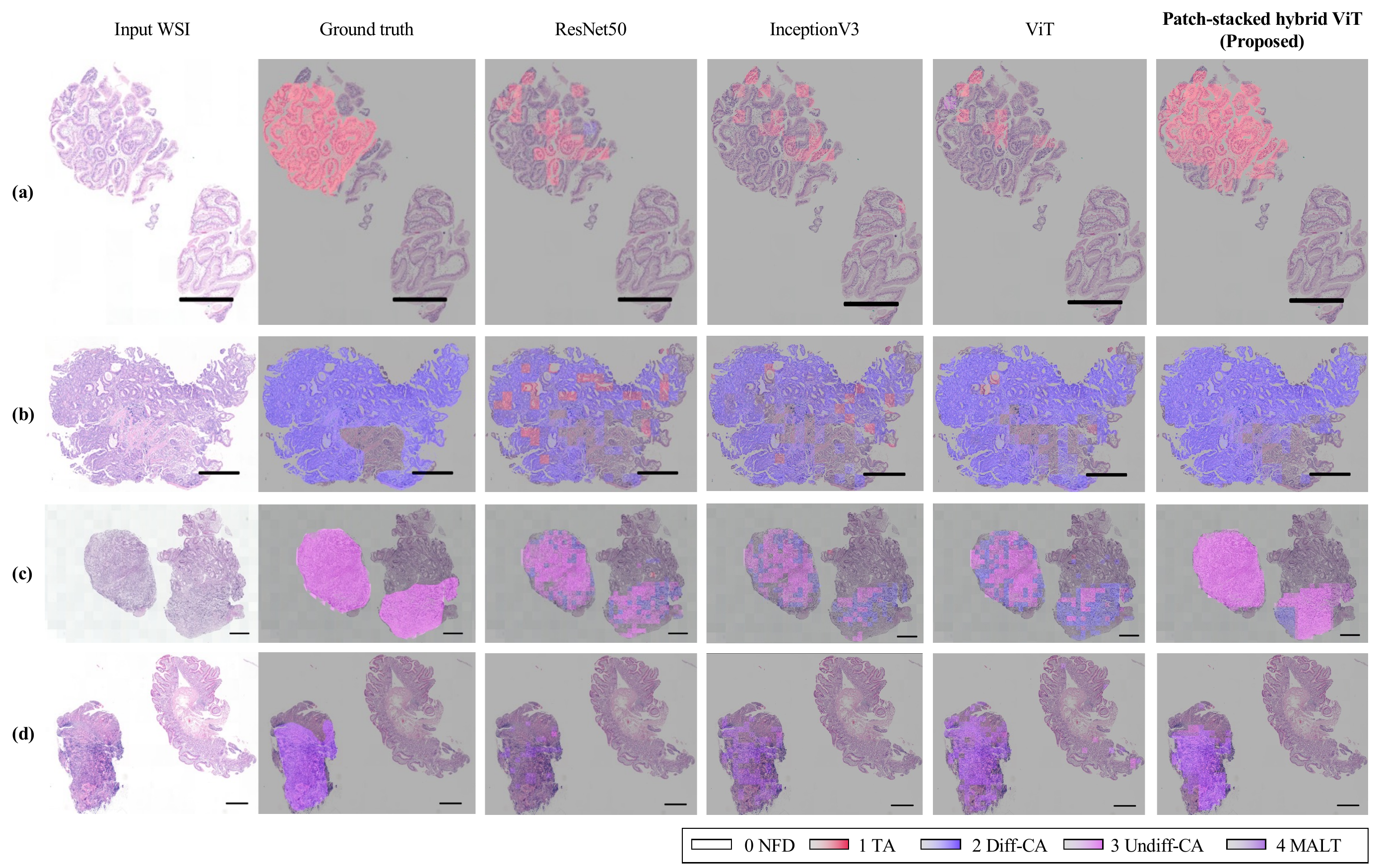}}  
		\caption{ROI-level probability map generation performance comparison. Representative cases selected from
			{\bf (a)} TA, {\bf (b)} Diff-CA, {\bf (c)} Undiff-CA and {\bf (d)} MALT classes. All the scale bars indicate 500  $\mu m$.}
		\label{fig:cnn_vit}
	\end{figure*}
\end{centering}

\section{Discussion}
\label{sec:discussion}

\subsection{Reliable Multi-class Classification Performance for Planning Gastric Cancer Treatment}

The AI system achieves promising diagnostic accuracy on the proposed five subclassifications, which cover almost all prevalent cases of external daily endoscopic cohort except for diffuse large B-cell lymphoma (DLBCL), as indicated as Others in Table \ref{Table:detailed_result}.  
\add{In this study, we exclude other types of lymphoma including DLBCL. Although DLBCL is one of the gastric malignancy, lymphomas are very rare (under 0.1\%), except the MALT lymphoma. 
We also exclude diseases such as gastrointestinal stromal tumor (GIST) or neuroendocrine tumor (NET) because these lesions usually presented as subepithelial masses that superficial gastric biopsy samples were not enough for pathology. 
We need to collect each subclass at least more than 3\% of entire dataset for training the AI system, but these diseases are not sufficient.
We rather focus on common tumorous diseases diagnosed from gastric biopsy including epithelial lesions and MALT lymphoma. Moreover, though H.pylori infection is important for the physician and clinical implementation for the treatment, H.pylori infection can be commonly diagnosed by other diagnostic tools, such as urea breath test, PCR test, or Giemsa staining. Herein, we exclude H.pylori infection and focus on planning cancer treatment on tumorous lesions. }

In fact, differentiating confusing adjacent classes, e.g., differentiated/undifferentiated and adenoma/adenocarcinoma, needs comprehensive understanding of structural characteristics. CNN-based patch-wise training has difficulty in understanding structural relationship between non-adjacent patches due to limited receptive field size. 
The proposed multiscale hybrid ViT models the long-range dependency among non-adjacent patch features after capturing short-range dependency among adjacent patch stacks exploited by CNN, thus the ViT-based framework shows promising multi-class classification performance compared to traditional CNN-based methods.
As shown in Fig. \ref{fig:cnn_vit}(c), the proposed patch-stacked hybrid ViT model provides stable ROI-level probability map with less confusing prediction between differentiated and undifferentiated tissues. Based on the improved ROI-level prediction results, the proposed network achieved diagnostic sensitivity of 1.00 for undifferentiated-type carcinoma class on all the test trials (see Table \ref{Table:metric}). 
Moreover, the proposed AI system's diagnostic sensitivity performance of MALT class demonstrates its capability for classifying MALT lymphoma and guiding clinician to proceed additional evaluation for the H.pylori infection.

\begin{centering}
	\begin{figure*}[h!]
		\centering
		\centerline{\includegraphics[width=0.96\linewidth]{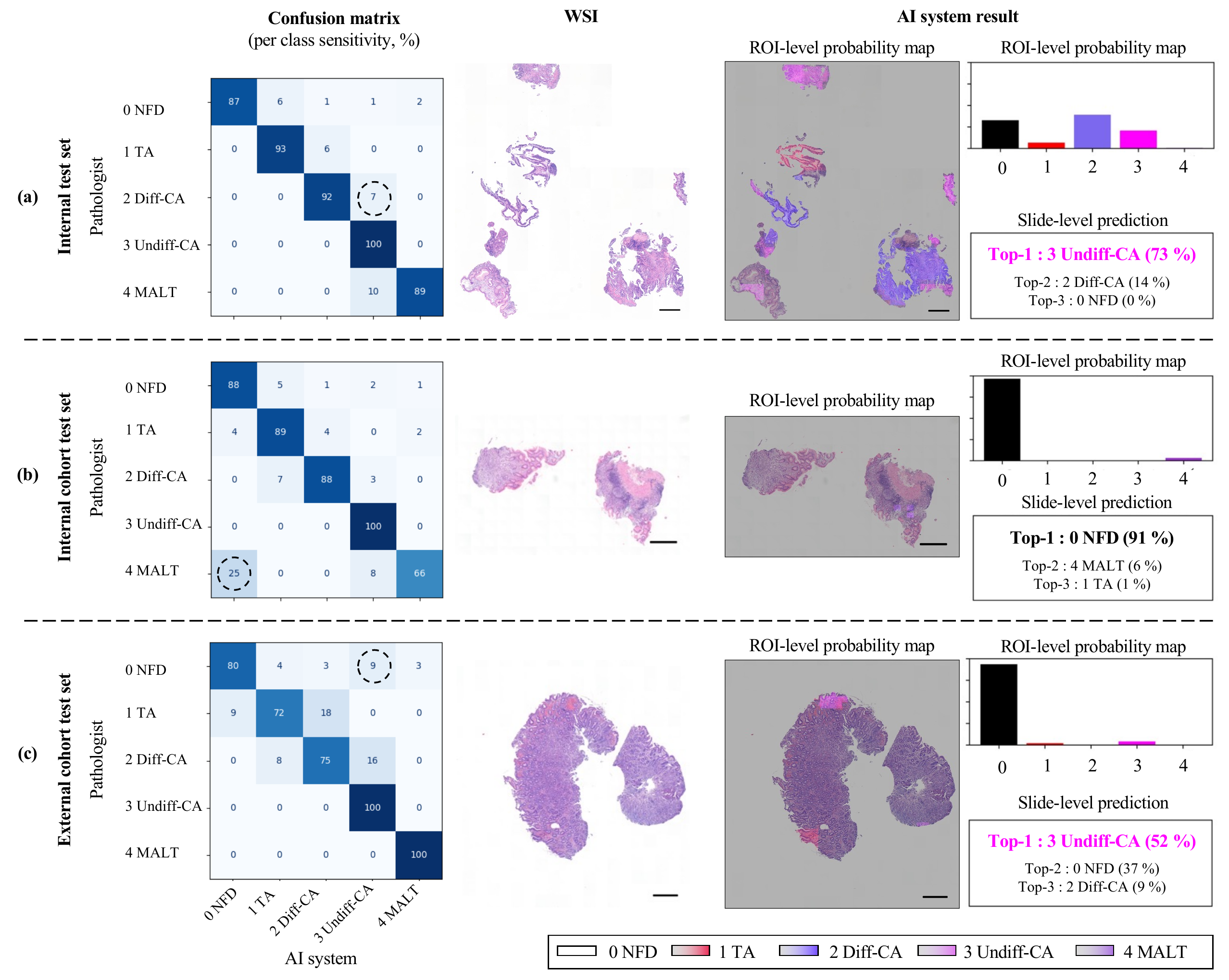}}
		\caption{Representative error cases. Falsely-diagnosed cases (indicated as dashed circles) selected from {\bf (a)} internal test set, {\bf (b)} internal cohort test set and {\bf (c)} external cohort test set. All the scale bars indicate 500 $\mu m$.}
		\label{fig:error}
	\end{figure*}
\end{centering}

\subsection{Limitations}
\label{sec:lim}

Our AI system is not free of limitations. As shown in Fig. \ref{fig:observer_roc}, the AI system ROC performance for TA, Diff-CA and Undiff-CA exceed average performance of human pathologists. However, for NFD and MALT, average human pathologist performance exceeds the AI system ROC performance. One reasonable solution for improving degraded class performance may be matching class-wise distribution between training data set and daily cohort test set. 
For example, NFD distribution (15\%) showed significant gap from that of the daily-acquired internal cohort test set (88\%), as shown in Table \ref{Table:dataset}. Similarly, daily internal cohort test sets contain around 30\% of MALT cases that require ancillary test, as analyzed in Fig. \ref{fig:observer}.
If our AI system was trained with large number of NFD samples or samples required ancillary test, it may have predicted samples of daily endoscopic screening dataset with reliable performance.

Secondly, variations in the staining condition may cause increased false-positive rate (18\%, 59 of 336 cases) of the external cohort test compared to that of the internal cohort test (10\%, 87 of 876 cases), as analyzed in Fig. \ref{fig:error}. This generalizability issue may be alleviated by developing advanced stain normalization techniques or training the system with multicentre datasets. 

\add{Lastly, our AI system architecture is limited in its receptive field. 
The slide-level network is designed to get multiple patches with dimension of 96 $\times$ 96 (total 9,216) patches by aggregating all the patch features throughout each endoscopic biopy sample. However, in order to adapt the proposed network to cover large tissue samples, almost 100M patch features need to be aggregated, as can be inferred from Table \ref{Table:tile}. To achieve this goal under a memory-efficient condition, Reisenbuchler et al. \cite{reisenbuchler2022local} proposed an efficient feature aggregation method by clustering unlimited patch features through k-nearest neighbor graphs. As a future study, we plan to expand our network to further analyze tissue samples by proposing an effective feature clustering method within the slide-level network.}

\add{Moreover, our ROI patch-level network is also limited in its input grid with 8 $\times$ 8 (total 64) patches for each patch stack, which is compromised depending on our computing resources. The restricted input gird dimension can degrade the patch-level network performance, especially when predicting ROI-level characteristics on the boundary of each endoscopic biopsy sample. For mitigating the performance degradation, we plan to design a memory-efficient ROI-level network structure without explicitly limiting its grid dimension for achieving better prediction accuracy.}

\begin{centering}
	\begin{table}[!h] \caption{\bf\footnotesize Comparison of patch feature aggregation method. \\}  \label{Table:tile}
		\centering
		\scalebox{0.7}{
			\begin{tabular}{c|l|cc|cc}
				\hline 
				\multirow{2}{*}{\textbf{Methods}} & \multirow{2}{*}{\textbf{Target sample}} &  \multicolumn{2}{c|}{\textbf{ROI patch-level network}} & \multicolumn{2}{c}{\textbf{Slide-level network}}   \\   
				~ & ~ & Aggregation & \#Patch & Aggregation & \#Patch \\ \hline  \hline				
				Ours & Endoscopic biopsy & ViT embedding   &  64  &  \multirow{4}{*}{ViT} & 9,216 \\ 
				\cite{wood2022enhancing} & Endoscopic biopsy &  RN-50 w/ K-Means  &  1 & ~ & 10,000  \\ 
				\cite{chen2022scaling} & Entire tissue  &  ViT embedding  &  256 & ~ & 104M  \\
				\cite{reisenbuchler2022local} & Entire tissue &  DN-101 w/ KNN  &  1 & ~ & Unlimited \\
				\hline

				\multicolumn{5}{l}{\textit{Note:} RN: ResNet; DN: DenseNet; KNN: k-nearest neighbor} \\
		\end{tabular}}
	\end{table}
\end{centering}

\subsection{Practical Usability in Clinical Settings}

Despite aforementioned limitations, our AI system has a strong point in providing explainable probability map together with the slide-level prediction. As shown in Fig. \ref{fig:cnn_vit}, pathologists can refer ROI-level probability maps to better understand the slide-level predictions, especially when the ROI-level probability distribution is non-dominant or the slide level probability is uncertain. Moreover, for region marked as suspicious for MALT class, as shown in Fig. \ref{fig:error}(b), pathologist can proceed ancillary test to confirm the final diagnosis of MALT lymphoma.



Finally, it should be noted that the results of the observer test show the stand-alone AI performance exceeds the average diagnostic performance of human pathologists for all classes.
Therefore, integrating  the AI system within the diagnostic workflow would be benefit of decreasing workloads on pathologists, while providing practical aid for planning surgical treatment, and can also provide diagnosis services for regions that have shortages in access to pathologists. 



\section{Conclusion}
\label{sec:conclusion}

Previous AI assistance systems were developed as a pre-analytic tool for early attention of the suspicious lesion in the cases and give second opinion to the pathologists. We proposed a multiscale hybrid ViT-based AI system capable of reliable multi-class classification, which can be matched to general gastric cancer treatment guidance. Assisted by the proposed AI system, clinicians can receive a presumptive pathologic opinion for predicting prognosis and planning appropriate cancer treatment.

\bibliographystyle{IEEEtran}
\bibliography{path2021}

\begin{thebibliography}{10}
\providecommand{\url}[1]{#1}
\csname url@samestyle\endcsname
\providecommand{\newblock}{\relax}
\providecommand{\bibinfo}[2]{#2}
\providecommand{\BIBentrySTDinterwordspacing}{\spaceskip=0pt\relax}
\providecommand{\BIBentryALTinterwordstretchfactor}{4}
\providecommand{\BIBentryALTinterwordspacing}{\spaceskip=\fontdimen2\font plus
\BIBentryALTinterwordstretchfactor\fontdimen3\font minus
  \fontdimen4\font\relax}
\providecommand{\BIBforeignlanguage}[2]{{%
\expandafter\ifx\csname l@#1\endcsname\relax
\typeout{** WARNING: IEEEtran.bst: No hyphenation pattern has been}%
\typeout{** loaded for the language `#1'. Using the pattern for}%
\typeout{** the default language instead.}%
\else
\language=\csname l@#1\endcsname
\fi
#2}}
\providecommand{\BIBdecl}{\relax}
\BIBdecl

\bibitem{sung2021global}
H.~Sung, J.~Ferlay, R.~L. Siegel, M.~Laversanne, I.~Soerjomataram, A.~Jemal,
  and F.~Bray, ``Global cancer statistics 2020: Globocan estimates of incidence
  and mortality worldwide for 36 cancers in 185 countries,'' \emph{CA: a cancer
  journal for clinicians}, vol.~71, no.~3, pp. 209--249, 2021.

\bibitem{jun2017effectiveness}
J.~K. Jun, K.~S. Choi, H.-Y. Lee, M.~Suh, B.~Park, S.~H. Song, K.~W. Jung,
  C.~W. Lee, I.~J. Choi, E.-C. Park \emph{et~al.}, ``Effectiveness of the
  korean national cancer screening program in reducing gastric cancer
  mortality,'' \emph{Gastroenterology}, vol. 152, no.~6, pp. 1319--1328, 2017.

\bibitem{jiang2020emerging}
Y.~Jiang, M.~Yang, S.~Wang, X.~Li, and Y.~Sun, ``Emerging role of deep
  learning-based artificial intelligence in tumor pathology,'' \emph{Cancer
  Communications}, vol.~40, no.~4, pp. 154--166, 2020.

\bibitem{bulten2020automated}
W.~Bulten, H.~Pinckaers, H.~van Boven, R.~Vink, T.~de~Bel, B.~van Ginneken,
  J.~van~der Laak, C.~Hulsbergen-van~de Kaa, and G.~Litjens, ``Automated
  deep-learning system for gleason grading of prostate cancer using biopsies: a
  diagnostic study,'' \emph{The Lancet Oncology}, vol.~21, no.~2, pp. 233--241,
  2020.

\bibitem{bejnordi2017diagnostic}
B.~E. Bejnordi, M.~Veta, P.~J. Van~Diest, B.~Van~Ginneken, N.~Karssemeijer,
  G.~Litjens, J.~A. Van Der~Laak, M.~Hermsen, Q.~F. Manson, M.~Balkenhol
  \emph{et~al.}, ``Diagnostic assessment of deep learning algorithms for
  detection of lymph node metastases in women with breast cancer,''
  \emph{Jama}, vol. 318, no.~22, pp. 2199--2210, 2017.

\bibitem{yoshida2021requirements}
H.~Yoshida and T.~Kiyuna, ``Requirements for implementation of artificial
  intelligence in the practice of gastrointestinal pathology,'' \emph{World
  Journal of Gastroenterology}, vol.~27, no.~21, p. 2818, 2021.

\bibitem{song2020clinically}
Z.~Song, S.~Zou, W.~Zhou, Y.~Huang, L.~Shao, J.~Yuan, X.~Gou, W.~Jin, Z.~Wang,
  X.~Chen \emph{et~al.}, ``Clinically applicable histopathological diagnosis
  system for gastric cancer detection using deep learning,'' \emph{Nature
  communications}, vol.~11, no.~1, pp. 1--9, 2020.

\bibitem{yoshida2018automated}
H.~Yoshida, T.~Shimazu, T.~Kiyuna, A.~Marugame, Y.~Yamashita, E.~Cosatto,
  H.~Taniguchi, S.~Sekine, and A.~Ochiai, ``Automated histological
  classification of whole-slide images of gastric biopsy specimens,''
  \emph{Gastric cancer}, vol.~21, no.~2, pp. 249--257, 2018.

\bibitem{park2021prospective}
J.~Park, B.~G. Jang, Y.~W. Kim, H.~Park, B.-h. Kim, M.~J. Kim, H.~Ko, J.~M.
  Gwak, E.~J. Lee, Y.~R. Chung \emph{et~al.}, ``A prospective validation and
  observer performance study of a deep learning algorithm for pathologic
  diagnosis of gastric tumors in endoscopic biopsies,'' \emph{Clinical Cancer
  Research}, vol.~27, no.~3, pp. 719--728, 2021.

\bibitem{iizuka2020deep}
O.~Iizuka, F.~Kanavati, K.~Kato, M.~Rambeau, K.~Arihiro, and M.~Tsuneki, ``Deep
  learning models for histopathological classification of gastric and colonic
  epithelial tumours,'' \emph{Scientific reports}, vol.~10, no.~1, pp. 1--11,
  2020.

\bibitem{jang2021deep}
H.-J. Jang, I.-H. Song, and S.-H. Lee, ``Deep learning for automatic
  subclassification of gastric carcinoma using whole-slide histopathology
  images,'' \emph{Cancers}, vol.~13, no.~15, p. 3811, 2021.

\bibitem{nagtegaal20202019}
I.~D. Nagtegaal, R.~D. Odze, D.~Klimstra, V.~Paradis, M.~Rugge, P.~Schirmacher,
  K.~M. Washington, F.~Carneiro, I.~A. Cree \emph{et~al.}, ``The 2019 who
  classification of tumours of the digestive system,'' \emph{Histopathology},
  vol.~76, no.~2, p. 182, 2020.

\bibitem{liu2017detecting}
Y.~Liu, K.~Gadepalli, M.~Norouzi, G.~E. Dahl, T.~Kohlberger, A.~Boyko,
  S.~Venugopalan, A.~Timofeev, P.~Q. Nelson, G.~S. Corrado \emph{et~al.},
  ``Detecting cancer metastases on gigapixel pathology images,'' \emph{arXiv
  preprint arXiv:1703.02442}, 2017.

\bibitem{dosovitskiy2020image}
A.~Dosovitskiy, L.~Beyer, A.~Kolesnikov, D.~Weissenborn, X.~Zhai,
  T.~Unterthiner, M.~Dehghani, M.~Minderer, G.~Heigold, S.~Gelly \emph{et~al.},
  ``An image is worth 16x16 words: Transformers for image recognition at
  scale,'' \emph{arXiv preprint arXiv:2010.11929}, 2020.

\bibitem{japanese2011japanese}
J.~G. C.~A. jgca@ koto. kpu-m.~ac. jp, ``Japanese classification of gastric
  carcinoma: 3rd english edition,'' \emph{Gastric cancer}, vol.~14, pp.
  101--112, 2011.

\bibitem{asenjo2007prevalence}
L.~Asenjo and J.~Gisbert, ``Prevalence of helicobacter pylori infection in
  gastric malt lymphoma: a sistematic review,'' \emph{Revista Espa{\~n}ola de
  Enfermedades Digestivas}, vol.~99, no.~7, p. 398, 2007.

\bibitem{yang2016management}
H.-J. Yang, S.~H. Lim, C.~Lee, J.~M. Choi, J.~I. Yang, S.~J. Chung, S.~H. Choi,
  J.~P. Im, S.~G. Kim, and J.~S. Kim, ``Management of suspicious
  mucosa-associated lymphoid tissue lymphoma in gastric biopsy specimens
  obtained during screening endoscopy,'' \emph{Journal of Korean Medical
  Science}, vol.~31, no.~7, pp. 1075--1081, 2016.

\bibitem{bacon2007mucosa}
C.~M. Bacon, M.-Q. Du, and A.~Dogan, ``Mucosa-associated lymphoid tissue (malt)
  lymphoma: a practical guide for pathologists,'' \emph{Journal of clinical
  pathology}, vol.~60, no.~4, pp. 361--372, 2007.

\bibitem{gotoda2000incidence}
T.~Gotoda, A.~Yanagisawa, M.~Sasako, H.~Ono, Y.~Nakanishi, T.~Shimoda, and
  Y.~Kato, ``Incidence of lymph node metastasis from early gastric cancer:
  estimation with a large number of cases at two large centers,'' \emph{Gastric
  cancer}, vol.~3, no.~4, pp. 219--225, 2000.

\bibitem{kook2019risk}
M.-C. Kook, ``Risk factors for lymph node metastasis in undifferentiated-type
  gastric carcinoma,'' \emph{Clinical endoscopy}, vol.~52, no.~1, p.~15, 2019.

\bibitem{gotoda2006endoscopic}
T.~Gotoda, H.~Yamamoto, and R.~M. Soetikno, ``Endoscopic submucosal dissection
  of early gastric cancer,'' \emph{Journal of gastroenterology}, vol.~41,
  no.~10, pp. 929--942, 2006.

\bibitem{khan2021transformers}
\BIBentryALTinterwordspacing
S.~Khan, M.~Naseer, M.~Hayat, S.~W. Zamir, F.~S. Khan, and M.~Shah,
  ``Transformers in vision: A survey,'' \emph{ACM Comput Surv}, 2021. [Online].
  Available: \url{https://doi.org/10.1145/3505244}
\BIBentrySTDinterwordspacing

\bibitem{he2016deep}
K.~He, X.~Zhang, S.~Ren, and J.~Sun, ``Deep residual learning for image
  recognition,'' in \emph{CVPR}, 2016, pp. 770--778.

\bibitem{grill2020bootstrap}
J.-B. Grill, F.~Strub, F.~Altch{\'e}, C.~Tallec, P.~Richemond, E.~Buchatskaya,
  C.~Doersch, B.~Avila~Pires, Z.~Guo, M.~Gheshlaghi~Azar \emph{et~al.},
  ``Bootstrap your own latent-a new approach to self-supervised learning,''
  \emph{Adv Neural Inf}, vol.~33, pp. 21\,271--21\,284, 2020.

\bibitem{abramoff2004image}
M.~D. Abr{\`a}moff, P.~J. Magalh{\~a}es, and S.~J. Ram, ``Image processing with
  imagej,'' \emph{Biophotonics international}, vol.~11, no.~7, pp. 36--42,
  2004.

\bibitem{wotherspoon1993regression}
A.~C. Wotherspoon, T.~Diss, L.~Pan, P.~Isaacson, C.~Doglioni, A.~Moschini, and
  M.~de~Boni, ``Regression of primary low-grade b-cell gastric lymphoma of
  mucosa-associated lymphoid tissue type after eradication of helicobacter
  pylori,'' \emph{The Lancet}, vol. 342, no. 8871, pp. 575--577, 1993.

\bibitem{lindley2014communicating}
S.~W. Lindley, E.~M. Gillies, and L.~A. Hassell, ``Communicating diagnostic
  uncertainty in surgical pathology reports: disparities between sender and
  receiver,'' \emph{Pathology-Research and Practice}, vol. 210, no.~10, pp.
  628--633, 2014.

\bibitem{scikit-learn}
F.~Pedregosa, G.~Varoquaux, A.~Gramfort, V.~Michel, B.~Thirion, O.~Grisel,
  M.~Blondel, P.~Prettenhofer, R.~Weiss, V.~Dubourg, J.~Vanderplas, A.~Passos,
  D.~Cournapeau, M.~Brucher, M.~Perrot, and E.~Duchesnay, ``Scikit-learn:
  Machine learning in {P}ython,'' \emph{J Mach Learn Res}, vol.~12, pp.
  2825--2830, 2011.

\bibitem{zhou2020deep}
S.~Zhou, H.~Marklund, O.~Blaha, M.~Desai, B.~Martin, D.~Bingham, G.~J. Berry,
  E.~Gomulia, A.~Y. Ng, and J.~Shen, ``Deep learning assistance for the
  histopathologic diagnosis of helicobacter pylori,'' \emph{Intelligence-Based
  Medicine}, vol.~1, p. 100004, 2020.

\bibitem{stegmuller2022scorenet}
T.~Stegm{\"u}ller, B.~Bozorgtabar, A.~Spahr, and J.-P. Thiran, ``Scorenet:
  Learning non-uniform attention and augmentation for transformer-based
  histopathological image classification,'' in \emph{Proceedings of the
  IEEE/CVF Winter Conference on Applications of Computer Vision}, 2023, pp.
  6170--6179.

\bibitem{reisenbuchler2022local}
D.~Reisenb{\"u}chler, S.~J. Wagner, M.~Boxberg, and T.~Peng, ``Local attention
  graph-based transformer for multi-target genetic alteration prediction,'' in
  \emph{Medical Image Computing and Computer Assisted Intervention--MICCAI
  2022: 25th International Conference, Singapore, September 18--22, 2022,
  Proceedings, Part II}.\hskip 1em plus 0.5em minus 0.4em\relax Springer, 2022,
  pp. 377--386.

\bibitem{wood2022enhancing}
R.~Wood, K.~Sirinukunwattana, E.~Domingo, A.~Sauer, M.~W. Lafarge, V.~H.
  Koelzer, T.~S. Maughan, and J.~Rittscher, ``Enhancing local context of
  histology features in vision transformers,'' in \emph{Artificial Intelligence
  over Infrared Images for Medical Applications and Medical Image Assisted
  Biomarker Discovery: First MICCAI Workshop, AIIIMA 2022, and First MICCAI
  Workshop, MIABID 2022, Held in Conjunction with MICCAI 2022, Singapore,
  September 18 and 22, 2022, Proceedings}.\hskip 1em plus 0.5em minus
  0.4em\relax Springer, 2022, pp. 154--163.

\bibitem{chen2022scaling}
R.~J. Chen, C.~Chen, Y.~Li, T.~Y. Chen, A.~D. Trister, R.~G. Krishnan, and
  F.~Mahmood, ``Scaling vision transformers to gigapixel images via
  hierarchical self-supervised learning,'' in \emph{Proceedings of the IEEE/CVF
  Conference on Computer Vision and Pattern Recognition}, 2022, pp.
  16\,144--16\,155.

\end{thebibliography}

%

\end{document}